\documentclass[11pt, a4paper]{article}
\usepackage[T1]{fontenc}
\usepackage{xcolor} 
\usepackage{comment}
\usepackage{graphicx}
\usepackage{authblk}

\begin{document}
\title{Interactive Evidence Maps for Visualizing and Understanding Systematic Reviews}
%
%
\author[1]{Aditi Mallavarapu\thanks{Corresponding author: amallav@ncsu.edu}}
\author[1]{Rohan Khandare}
\author[1]{Mokshagna Kadiyala}
\author[1]{Neelesh Yaddanapudi}
\author[2]{Noah L. Schroeder}
\author[2]{Shan Zhang}
\author[3]{Jessica R. Gladstone}

\affil[1]{North Carolina State University, Raleigh, NC 27695}
\affil[2]{University of Florida, Gainesville, FL, 32611}
\affil[3]{University of Illinois at Urbana-Champaign, Champaign, IL, 61820}
\maketitle              
\begin{abstract}
Systematic reviews provide comprehensive syntheses of research fields. As a result, systematic reviews often emphasize synthesizing across the large bodies of literature rather than just describing the studies from which the conclusions were drawn. This risks an incomplete description of the sample - encouraging overgeneralization of the findings, obscuring connections between existing work, or overshadowing gaps in the literature. To address this challenge, we introduce interactive evidence maps; an accessible visualization tool that enables researchers to explore, filter, and analyze review data dynamically. Our approach leverages large language models to extract topic models that structure heterogeneous review data into an interactive, explorable knowledge map that supports deeper inspection beyond static tables and figures. We demonstrate the usefulness of interactive evidence maps using data from a published scoping review of pedagogical agents in K-12 education, and compare the results of the evidence map to those reported in the scoping review. Results show that interactive evidence maps complement traditional syntheses by enhancing transparency, supporting exploratory analysis, and revealing patterns and gaps that may not be easy to detect through narrative summaries alone. 

\end{abstract}
\section{Introduction}
Systematic reviews and meta-analyses serve as cornerstone methodologies in educational research. They have been providing comprehensive syntheses that inform evidence-based practice and policy decisions \cite{newman_systematic_2020}. In the field of Artificial Intelligence in Education (AIED) as in others, systematic reviews have been instrumental in mapping the landscape of learning technologies, pedagogical interventions, and their effectiveness across diverse educational contexts \cite{chen_artificial_2020}. However, the very comprehensiveness that makes systematic reviews valuable also presents significant challenges. As review authors synthesize findings across large bodies of literature, the emphasis often shifts toward high-level, synthesized conclusions that are "broad strokes" at the expense of detailed study-level descriptions. This synthesis-focused approach, while necessary for generating actionable insights, risks encouraging overgeneralization of findings, obscuring nuanced connections between studies, and overshadowing critical gaps in the literature that warrant further investigation.

The challenge of balancing synthesis with transparency has become increasingly acute as the volume of research continues to grow exponentially. Traditional presentation formats—static tables, narrative summaries, and fixed visualizations — all struggle to accommodate the complexity and heterogeneity inherent in modern systematic reviews \cite{martin_synthesis_2020} regardless of the research field. Fixed visualizations, such as predefined bar charts, forest plots, and summary graphs embedded in review documents, offer visual clarity but lack flexibility for reader-driven exploration \cite{keim2008visual,vieira2018visual}. For example, a systematic review might include a bar chart displaying the number of studies conducted at different educational levels (elementary, middle, high school), providing a useful overview of the evidence distribution. However, this visualization is fixed: readers cannot interact with it to answer follow-up questions such as "Which of these elementary school studies focused on mathematics?" or "How many randomized controlled trials were conducted at the high school level?" \cite{sedlmair2012design,kossmeier_charting_2020}. Similarly, a scatter plot showing publication trends over time cannot be filtered to examine trends within specific intervention types or to isolate studies with particular outcome measures. These static representations reflect the review authors' analytical priorities but cannot adapt to the diverse questions and interests that different readers bring to the evidence base \cite{card2009information}. The limitations of static visualizations in systematic reviews have been noted by multiple authors, who emphasize the need for more flexible, interactive approaches that allow users to explore evidence according to their own analytical needs \cite{thomas2017methods,beller2018interactive,kossmeier_charting_2020,owen_metainsight_2019,schwendimann2017perceiving,vieira2018visual}. Readers are often left unable to interrogate the underlying evidence base, explore subsets of studies relevant to their specific interests, or identify patterns that may not align with the authors' primary synthesis narrative. This limitation is particularly problematic in interdisciplinary fields like AIED, where research spans multiple methodologies, theoretical frameworks, and application contexts. While existing visualization tools such as Open Knowledge Maps \cite{kraker2016open}, VOSviewer \cite{vaneck2010software}, and CmapTools \cite{canas2004cmap} have advanced bibliometric and knowledge mapping capabilities, they often fall short in supporting the specific analytical needs of systematic review synthesis, particularly in enabling dynamic exploration of study characteristics, outcomes, and contextual factors that define educational technology research \cite{abu-salih_systematic_2024}.

To address this gap, we introduce interactive evidence maps — a novel approach that leverages large language models (LLMs)- based topic modeling to transform systematic review data into dynamic, explorable knowledge structures. Our approach combines automated topic extraction with interactive visualization to create evidence maps that enable researchers, practitioners, and policymakers to engage with review findings at multiple levels of granularity. Unlike static synthesis documents, interactive evidence maps support filtering, clustering, and exploratory analysis of the underlying study corpus, revealing patterns and gaps that may remain hidden in traditional narrative summaries. By maintaining the connection between high-level synthesis and study-level detail, our approach enhances transparency, supports diverse user needs, and complements rather than replaces traditional systematic review methodologies.

\section{Related works}
\subsection{Visualization and Knowledge Mapping in Research Synthesis}

The visualization of research landscapes has a rich history across multiple disciplines, with various tools and methodologies developed to support bibliometric analysis \cite{borner2003visualizing,borner2010atlas} and knowledge mapping \cite{eppler2008knowledge,canas2004cmap}. These contributions establish important principles for effective scientific publication-derived visualizations, including the importance of interactivity, multi-scale representation, and semantic clarity. However, while these frameworks excel at mapping entire scientific disciplines and citation networks at scale, they have been typically applied to comprehensive bibliometric databases rather than the curated, systematically reviewed evidence sets that characterize systematic reviews in specific research questions.

\textit{VOSviewer} \cite{vaneck2010software}, and its extension \textit{CitNetExplorer} \cite{vaneck2014visualizing}, have been popular for constructing and visualizing bibliometric networks. These focuses on citation-based relationships and keyword co-occurrence, with limited support for incorporating study-level characteristics, methodological features, or outcome data that are central to systematic reviews. Web-based mapping tools like \textit{Open Knowledge Maps} \cite{kraker2016open}, LitMaps \cite{litmaps2024} and OpenAlex \cite{priem2022openalex} combine text mining with visualization to create accessible overviews of research topics.  However, the automated, black-box approach that extracts this information lacks the systematic rigor and the comprehensive inclusion criteria that are core to systematic reviews. Firstly, these tools cannot easily incorporate the detailed coding and quality assessments that systematic reviewers apply to included studies. Additionally, these can only reveal broad strokes of literature review and require technical expertise to extract comprehensive synthesis information, limiting their accessibility for our use-case.

\subsection{Topic Modeling and Computational Approaches for Synthesis and Knowledge Mapping}
Computational approaches to understanding research landscapes have evolved significantly with advances in natural language processing and machine learning. Traditional topic modeling approaches, particularly Latent Dirichlet Allocation (LDA), have been applied to a large corpora of scientific literature to identify latent thematic structures \cite{blei2003latent}. While these methods can reveal topic-driven patterns across thousands of documents, they often produce topics that lack semantic coherence or that do not align with meaningful distinctions in systematic review contexts.

BERTopic and its scientific variants (e.g, SciBERT \cite{beltagy_scibert_2019}, Specter2 \cite{singh_scirepeval_2023}) represents a more recent advancement in topic modeling post their not very sophisticated semantic predecessor LSA (Latent Semantic Analysis) \cite{deerwester1990indexing}. These transformer-based approaches generate more semantically coherent and contextually relevant topic clusters \cite{grootendorst2022bertopic}. Simply put, they combine document embeddings with class-based TF-IDF (Term Frequency-Inverse Document Frequency) or other similarity metric representations to produce topics by capturing semantic nuance and distinguishing key terms. These have been known to generate more accurate representations and cluster differentiation that better reflect human interpretations of thematic structures than their predecessors. To date, BERTopic and its decedents have been sparsely used for systematic review contexts (for an example, \cite{zhang2024semi}). These limited instances have  
been influential in showcasing the value in integrating structured coding schemes with topic modeling for systematic analysis. However, the approaches to present these results in a usable manner that can facilitate this needs to be explored.


One approach that has been popular is to integrate the topic modeling approaches with clustering algorithms and dimensionality reduction techniques to organize research literature as knowledge maps \cite{marrone_interdisciplinary_2020,cobo2011science}. While topic models can lead to uninterpretable information, knowledge maps can be represented as visual diagrams that afford mapping, and navigation, across multiple layers of information within the evidence. In addition to revealing topics/subtopics - these techniques can reveal proximity relationships between studies and if integrated with human coded information can reveal specific facets that could facilitate a multi-level synthesis of systematic reviews. 

\subsection{Visual Approaches to Systematic Review}
While visualization tools are common in bibliometrics, their specific application to systematic review synthesis remains relatively underdeveloped. Traditional systematic reviews rely primarily on tables and forest plots for meta-analyses \cite{lewis2001forest}, and narrative text to present findings. Forest plots effectively communicate effect sizes and confidence intervals but are limited to quantitative studies and do not capture the broader characteristics of the evidence base \cite{lewis2001forest}. Evidence and gap maps represent one approach to visualizing the scope and distribution of evidence within systematic reviews \cite{snilstveit2016evidence}. These maps typically use matrix structures to show which interventions have been studied in which populations or contexts, with cells indicating the volume and sometimes the quality of available evidence. While useful for identifying gaps, traditional evidence and gap maps are static and do not support dynamic exploration or filtering based on multiple study characteristics simultaneously.

The application of network visualization to systematic reviews is not a novel concept. Radianti et al. \cite{radianti2021systematic} conducted a systematic review employing network visualization techniques to map the relationships between studies, themes, and application domains. Their approach demonstrated how visualization can enhance the presentation of review findings, making patterns and clusters within the evidence base more apparent. Other studies have incorporated visual analytics to present findings more effectively, in the form of heatmaps to show patterns of research activity across different dimensions, sankey diagrams to illustrate participant flow or intervention pathways \cite{glover_sankey_2022,haddaway_eviatlas_2019}, and timeline visualizations to show temporal trends. However, these implementation remained primarily descriptive and did not fully integrate interactive exploration capabilities that would allow users to dynamically filter and analyze the evidence base according to their specific interests. 

\subsection{Contribution}
Despite these advances in visualization tools, topic modeling, and visual systematic review methodologies, a significant gap remains in understanding effective strategies that combine the rigor of systematic reviews with the exploratory power of interactive knowledge maps. 
This paper provides a proof-of-concept interactive evidence mapping tool exploring strategies that combine: (1) the systematic rigor of traditional reviews, (2) the pattern-discovery capabilities of topic modeling powered by LLMs, and (3) the exploratory flexibility of interactive visualization. Our approach transforms systematically coded review data into dynamic knowledge maps that support multiple levels of analysis — from high-level thematic overviews to detailed inspection of individual studies. By maintaining explicit connections between automated topic models and underlying study characteristics, our evidence maps enhance transparency, support diverse user needs, and reveal patterns that may not emerge from static synthesis narratives alone. We demonstrate this using data from a published scoping review of pedagogical agents in K-12 education \cite{zhang2025pedagogical} comparing insights from the interactive evidence map with those reported in the original review to illustrate both the complementary value and the new analytical possibilities that interactive evidence mapping provides.

\begin{figure}
    \centering
    \includegraphics[width=.8\linewidth]{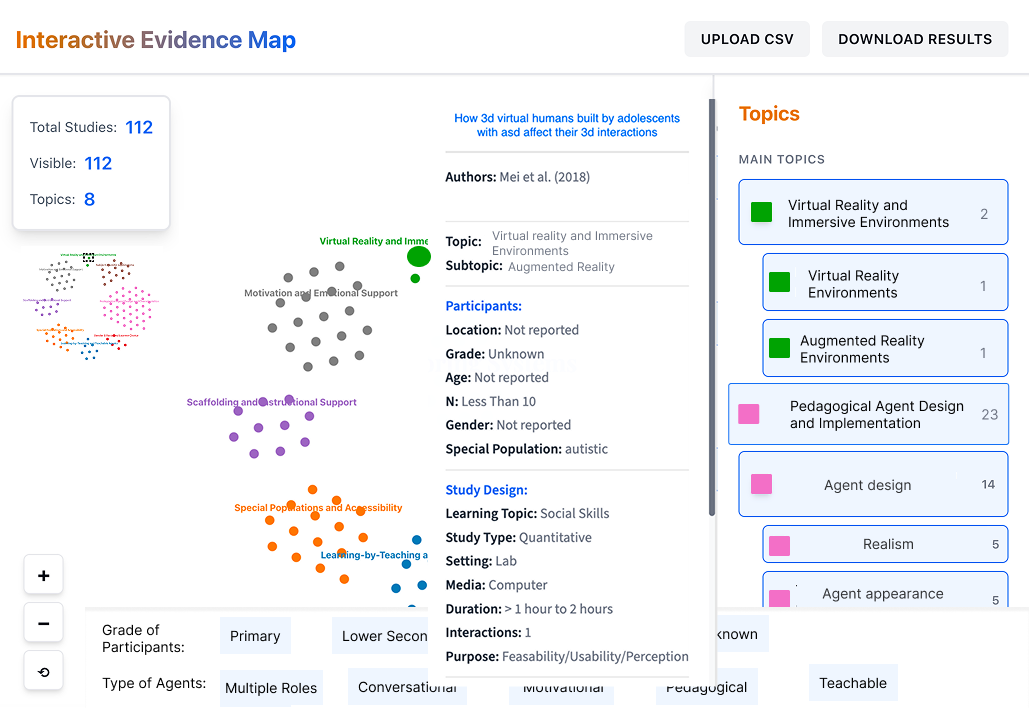}
    \caption{Proof-of-concept implementation of the evidence map tool with a information and navigation tile (top left). The tool visualizes the papers as nodes with colors representing modeled topics, allows exploration of modeled topics and sub-topics (right panel), selection of coded items (bottom panel) and affords navigation of individual papers through detail panel (center).} 
    \label{fig:fullscreen}
\end{figure}

\section{Pilot Implementation}
For the generation of interactive evidence maps our LLM-integrated pipeline transforms systematically coded review data into an explorable visual knowledge structure (Figure \ref{fig:fullscreen}). The process begins with \textit{data preparation}, where systematically reviewed studies—already extracted and coded according to a predefined protocol, are uploaded to the pipeline in a structured CSV format. Each row in the file represents a study, with columns capturing key characteristics such as bibliographic information, and systematic review specific factors like details of study design, participant demographics, intervention, outcomes, and any qualitative findings or themes identified by the original reviewers. Everything except bibliographic information are flexible to be customized as per the reviewers' findings and domain practices. 

In the \textit{topic modeling} stage, we employ a prompt-based approach using Claude (Anthropic's LLM) \cite{Anthropic2024Claude} that is freely accessible (as of Jan, 2026) through the API to extract thematic clusters from the review data.  Our prompts provide Claude with the structured study data and requests identification of coherent topics based on study abstracts and titles. The prompt is simple, \textit{``Analyze this \textit{dataset} research studies and identify 6-8 main topics and 2-3 subtopics for each.''}. Here the word dataset is replaced by the coded data. Claude returns topic labels, descriptions and assigns each study to one topic based on semantic similarity. These assignments and descriptions are available for download. We used a prompt-based approach as it allows for rapid prototyping and iterative refinement for this initial proof-of-concept, though future iterations will implement pretrained yet fine-tunable API-based models that enable automated, scalable and transparent topic (to the extent possible) extraction for larger scientific datasets. 

In the \textit{data structuring} phase the topic assignments are integrated with the original coded study characteristics, creating a unified data structure that links each study to its topics while preserving all systematically extracted metadata. This structured data is then transformed into a format compatible with interactive visualization libraries. In the visualization generation stage, we employ D3.js and React-based components to create the interactive evidence map interface. The visualization includes multiple coordinated views (see figure \ref{fig:fullscreen}): (1) (center) a spatial layout showing topics clusters with nodes for each study classified under that topic, the size of the cluster is proportional to the number of studies, (2) filtering controls that allow users to subset studies by topics (right) and another panel to filter studies by coded features such as publication year, methodology, educational level, or outcome type (bottom), (4) detail panels that display study-level information when users hover over or select individual studies (center right), (5) dynamic statistics that update based on applied filters (top-left corner) and (6) navigation map to guide spatial exploration for large maps.

Our initial pilot implementation code will be made available on github. 

\section{Proof-of-concept: Evidence Map for Guiding Synthesis of Systematic Reviews}

\subsection{Dataset}
We used the 112 studies that were included in \cite{zhang2025pedagogical}'s scoping review as our dataset. The review coded factors including learning topics (e.g., science, mathematics, social skills, language), agent types (Pedagogical, Conversational, Multiple roles), grade levels (primary, lower secondary, upper secondary), agent roles (coaching/scaffolding, information source), study purpose (experiment, feasibility/usability), among many others. 

Although the dataset consists of 112 studies and the authors provided the entire coding form complete with data from each study, identifying which intersectional factors were popular and where research gaps existed is a challenge. This would require users to manipulate multiple column data in a CSV. Questions like ``in which domains have conversational agents been used for assessments?'' or ``how many studies were included for coaching pedagogical agent in comparison to ones used as information source''  would be very challenging to investigate in this current form. 

\subsection{Topic Model}
As shown in Figure \ref{fig:fullscreen}, the interactive evidence map starts at the furthest out view, showing eight defined topics for this dataset. Despite the prompt being very generic, it is worth noting that the generated topics reflected meaningful distinctions in the evidence base such as methodological approaches (e.g., learning by teaching paradigms), technological implementations (e.g., virtual reality and immersive environments, pedagogical agent design and implementation), pedagogical and theoretical frameworks (e.g., scaffolding and instructional support, learning outcomes and achievement, motivation and emotional support), application areas (e.g., social perception and human agent interaction) and contextual factors (e.g., individual differences and learner characteristics). While the modeled topics are not exactly orthogonal to dimensions coded by the systematic review team, they had an overlap with some of the coded information.

\subsection{Data exploration}
On the right side of Figure \ref{fig:fullscreen}, topic filters help narrow down the sample along with the coded features in the bottom panel. The tool derives the topic and subtopic labels from the topic model and displays number of studies in each category to guide quantitative comparison across topics and subtopics. The bottom panel is populated using the column headings from coded study details. 

One can zoom in on any of the clusters, down to the level of the specific paper within the cluster (not currently shown in Figure \ref{fig:fullscreen}),  and click on a node to read details of an individual study (center right in Figure \ref{fig:fullscreen}). These workflows enable the users to explore how the extracted topics and coded information intersect, complementing the human review process. Moreover, it allows the reader to see the dimensions on which the studies cluster together, and guide the identification of specific papers of interest to them based on the topics and coded factors. Asking questions like: ``which studies discussed learning by teaching agents for primary grade students?'' affords the user four layers of integrated information. First, the user could select \textit{learning by teaching} in the topic filter (right panel) and primary option in the feature filter panel (bottom). The panel buttons become active or inactive based on how the studies were coded. The user can count the studies and know other intersecting details like when these studies were published and what settings (e.g., classroom settings) the studies employed. Further they can zoom-in to see the sub-topics and know details about each study highlighted. The navigation panel can guide the users in case the evidence maps are large. Additionally if certain papers were labeled with multiple topics - the detail panel would display this information.

\subsection{Synthesizing Findings for Pedagogical Agents for K-12 Education}
The interactive evidence map enabled multi-layered pattern discovery that transcends what either thematic clustering or human coding reveals in isolation.

\textit{Uncovering Hidden Design Principles}. When exploring the "Special Populations and Inclusive Design" cluster, three sub-themes were identified: perception studies, social appearance investigations, and human-AI interaction research. While the first two subtopics contained numerous studies comparing design variations, patterns visible in human coding, the less populated human-AI interaction subtopic revealed a coherent design philosophy invisible in the coded data  or in the subtopic alone: studies addressing students with Autism Spectrum Disorder consistently employed non-human agent forms focused on social skills development, regardless of subject matter. This cross-cutting principle only emerged when clustering isolated studies by their content rather than just the pre-defined categories.

\textit{Revealing Temporal Evolution}. Cross-filtering topics with human-coded temporal and methodological dimensions exposed three distinct research epochs. The first epoch (2003-2010) prioritized efficacy validation through laboratory experiments asking "does it work?". The second epoch (2011-2018) shifted toward design optimization, systematically varying agent form and role — particularly conversational agents for social skills — building on earlier proof-of-concept successes. The more recent epoch (2019-2023) emphasizes ecological validity through classroom-based, longitudinal deployments with specific populations, tracking motivational outcomes as learners interact with agents posing as coaches over extended periods. This progression from controlled experiments to naturalistic implementation mirrors broader technological maturation in this research vein.

\textit{Exposing Critical Gaps}. The dual-filtering system highlighted conspicuous absences: only three studies examined conversational agents with upper secondary learners; language learning interventions for adolescents (ages 10-19) were entirely absent; and non-human agent designs appeared exclusively in Autism-focused social skills interventions, suggesting missed opportunities for applying these design patterns to neurotypical populations or other learning domains. Conversely, the map revealed saturation in primary-level science education with pedagogical agents in classroom settings, signaling diminishing returns for incremental contributions in this configuration while underexplored combinations (e.g., secondary-level language learning, adolescent-focused conversational systems) remain ripe for investigation. By enabling researchers to fluidly toggle between content-driven thematic clusters and methodological filters, the tool transforms systematic reviews from static summaries into analytical workspaces where cross-dimensional patterns, temporal trajectories, and strategic research gaps become immediately visible through interactive exploration rather than laborious manual synthesis.

\section{Discussion and Next Steps}
The dual-filtering architecture we implemented, where LLM generated topics derived from titles and abstracts reveal emergent thematic patterns while human-coded categories enable filtering by methodological dimensions, addresses a key limitation of traditional systematic reviews. 
However, several open questions warrant further investigation: 
\begin{enumerate}
    \item \textit{Scalability and performance} with evidence bases exceeding 500-1000 studies, where both topic model coherence and visualization responsiveness require empirical validation.
    \item \textit{Text granularity} trade-offs between abstract-based clustering (limited context but universally available) versus full-text analysis (richer semantic information but requiring additional data preparation and computational resources).
    \item \textit{Probabilistic assignment transparency}, as the current implementation assigns each study to a single topic-subtopic pair without exposing confidence scores, alternative candidate topics, or co-occurrence patterns. Future iterations should visualize assignment uncertainty and enable multi-label views where studies legitimately span multiple thematic clusters.
    \item \textit{LLM output stability and reproducibility}, a critical concern for systematic reviews which demand replicable methods. The black-box behavior of prompt-based large language models means that cluster assignments and labels can vary substantially across runs and stochastic sampling during generation potentially undermines the reproducibility standards expected in evidence synthesis. \item \textit{Cross-domain generalizability}, testing whether the approach transfers effectively to systematic reviews in fields with different coding conventions (e.g., medical interventions, policy evaluations, engineering solutions). 
\end{enumerate}

Moving forward, we envision an open-source platform with an API-based architecture supporting pluggable topic models (LLMs with temperature controls and seed parameters for partial stability, alongside deterministic alternatives like SciBERT and Specter2), model comparison metrics that quantify cluster stability across multiple runs, and customizable visualization layouts, enabling researchers to upload coded data, configure analysis parameters through a graphical interface, compare LLM outputs against traditional topic modeling baselines, and generate publishable interactive evidence maps. In sum, we envision helping researchers transform their static CSV tables into explorable research landscapes that balance the discovery potential of AI-driven clustering with the reliability requirements of systematic review methodology.

\section{Conclusion}
Systematic reviews tend to provide a high level synthesis in order to draw generalizable conclusions. In order to accomplish this it is easy to accidentally overgeneralize results or provide limited context about the descriptive characteristics of the sample, and there is a documented need for new ways to visualize data according to a reader's needs \cite{thomas2017methods,beller2018interactive,kossmeier_charting_2020,owen_metainsight_2019,schwendimann2017perceiving,vieira2018visual}. Our proof-of-concept interactive evidence maps based on language models and knowledge mapping technique highlighted the clear benefits with regards to exploring the characteristics of the sample. We believe that interactive evidence maps provide a promising opportunity to be reported alongside traditional systematic review reporting schemes. 

\section*{Funding} 
Authors Mokshagna Kadiyala and Neelesh Yaddanapudi were funded by REUs from the NC State Engineer Your Experience program. This material is based upon work supported by the National Science Foundation under Grant DUE-2518159 and work supported by the National Science Foundation and the Institute of Education Sciences under Grant DRL-2229612. Any opinions, findings, and conclusions or recommendations expressed in this material are those of the author(s) and do not necessarily reflect the views of the National Science Foundation or the U.S. Department of Education.

%
%
%
 \bibliographystyle{unsrt}
 \bibliography{mybibliography}

\end{document}